\documentclass[%
aip,
apl,
reprint,
]{revtex4-1}
\usepackage[english]{babel}
\usepackage{amssymb,amsmath,stmaryrd,array}

\usepackage{graphicx}

\makeatletter

\newcommand{\dega}{$^\circ$}
\newcommand{\degc}{$^\circ$C}
\newcommand{\myarrow}[1]{\ensuremath{\xrightarrow{{#1}^\circ\mathrm{C}}}}

 \newcommand{\vc}[1]{\mathbf{#1}}
 
 \newcommand{\uvc}[1]{\hat{\mathbf{#1}}}
 
 \newcommand{\ind}[1]{\mathrm{#1}}

\makeatother

\begin{document}
\DeclareGraphicsExtensions{.eps,.png,.pdf}
\title{%
Enhanced Kerr effect in vertically aligned deformed helix ferroelectric
liquid crystals
}

\author{E.P.~Pozhidaev}
\affiliation{%
P.N. Lebedev Physical Institute of the Russian Academy of Sciences, 
Leninskiy pr. 53, Moscow, 119991 Russia
}
\affiliation{%
Department of Electronic and Computer Engineering, 
Hong Kong University of Science and Technology,
Clear Water Bay, Kowloon,
Hong Kong}

\author{A. K.~Srivastava}
\affiliation{%
Department of Electronic and Computer Engineering, 
Hong Kong University of Science and Technology,
Clear Water Bay, Kowloon,
Hong Kong}

\author{Alexei~D.~Kiselev}
\email[Email address: ]{kiselev@iop.kiev.ua}
\affiliation{%
 Institute of Physics, National Academy of Sciences of Ukraine, 
prospekt Nauki 46, 03680 Kiev, Ukraine}
\affiliation{%
Department of Electronic and Computer Engineering, 
Hong Kong University of Science and Technology, 
Clear Water Bay, Kowloon, Hong Kong}

\author{V.G.~Chigrinov}
\email[Email address: ]{eechigr@ust.hk}
\affiliation{%
Department of Electronic and Computer Engineering, 
Hong Kong University of Science and Technology,
Clear Water Bay, Kowloon,
Hong Kong}

\author{V.V.~Vashchenko}
\affiliation{%
State Scientific Institution ``Institute for Single Crystals'', Lenin
Ave., 60, Kharkov 61001, Ukraine
}

\author{A.I. Krivoshey}
\affiliation{%
State Scientific Institution ``Institute for Single Crystals'', Lenin
Ave., 60, Kharkov 61001, Ukraine
}

\author{M.V.~Minchenko} 
\affiliation{%
P.N. Lebedev Physical Institute of the Russian Academy of Sciences,
  Leninskiy pr. 53, Moscow, 119991 Russia }

\author{H.-S.~Kwok}
\affiliation{%
Department of Electronic and Computer Engineering, 
Hong Kong University of Science and Technology,
Clear Water Bay, Kowloon,
Hong Kong}

\date{\today}

\begin{abstract}
We disclose the vertically aligned deformed helix
ferroelectric liquid crystal (VADHFLC) whose Kerr constant
($K_{\mathrm{kerr}}\approx 130$~nm/V$^2$ at $\lambda=543$~nm) is one
order of magnitude higher than any other value previously reported for
liquid crystalline structures.
Under certain conditions, the phase modulation
with ellipticity less than $0.05$ over the range of continuous and hysteresis free electric adjustment of
the phase shift from zero to 2$\pi$ have been obtained at sub-kilohertz frequency.
\end{abstract}

 \maketitle


Pure phase modulation of light with conserved
ellipticity  is in high demand for a variety of
applications. These include photonic devices such as
tunable lenses, focusers, wave front correctors and
correlators~\cite{Naumov:optex:1999,Ren:apl:2003,Xu:optex:2012,Hu:optex:2004} 
used as building blocks of optical
information processors and displays. Nowadays
microelectromechanical
systems~\cite{Rogg:oe:1997},
micro-opto-electromechanical systems~\cite{Mota:oe:1997} 
and bimorph deformable mirrors~\cite{Adelman:aplopt:1977}
are employed for the high frequency ($f>1$~kHz)
binary phase modulation with 
the fixed  phase shift arising due to light reflection. However, the
progress in the development of optical processing systems
is being impeded by the lack of high performance
and high-speed liquid crystal (LC) light phase modulators
with continuous and hysteresis free response.~\cite{Shraug:spie:2001}

Phase modulation of light
based on the Kerr effect in
polymer stabilized blue phase liquid
crystals (PSBPLC) was recently explored in~\cite{Hisakado:advmat:2005,Chen:apl:2013}.
The largest Kerr constant, $K_{\mathrm{kerr}}$, reported
for PSBPLCs at the wavelength $\lambda=514$~nm  is
33.1~nm/V$^2$.~\cite{Chen:apl:2013}
  The electro-optical (EO)
response time for such systems is limited to the
millisecond range. 
Moreover, 
a pronounced EO hysteresis
caused by the polymer network present in 
PSBPLC systems 
is not appropriate for applications.

An alternative approach deals with short pitch
cholesteric liquid crystals (ChLCs).~\cite{Lee:jap:2009}
The polymer stabilized standing helix ChLC 
based on in-plane addressing~\cite{Gardiner:apl:2011} 
shows the response times $\approx 50$~$\mu$s. 
But, for the 2$\pi$ phase modulation, 
very high electric fields, $E\approx 10$~V/$\mu$m, 
are required. 

In Ref.~\cite{Kiselev:pre:2013}, we found that the orientational Kerr effect
in a vertically aligned deformed helix
ferroelectric LC (VADHFLC)
with subwavelength helix pitch, $p_0\approx150$~nm,
is characterized by fast 
and, under certain conditions,~\cite{Blinov:pre:2005,Pozhidaev:jsid:2012}
hysteresis-free electro-optics.
A typical
EO response time is around 100~$\mu$s and is 
almost independent of applied electric field. 
Though the Kerr constant~\cite{Kiselev:pre:2013} 
$K_{\mathrm{kerr}}\approx 27$~nm/V$^2$
at $\lambda=543$ nm
is already comparable with  $K_{\mathrm{kerr}}$ of 
the best modern PSBPLC,~\cite{Chen:apl:2013}
it is feasible to increase $K_{\mathrm{kerr}}$
of VADHFLCs further,
which is of vital importance for low voltage phase modulators
of light. 
In this letter we suggest an approach 
to enhance the Kerr effect and use it
to drastically increase 
the Kerr constant of VADHFLC.

\begin{figure}[!tbh]
\centering
\resizebox{85mm}{!}{\includegraphics*{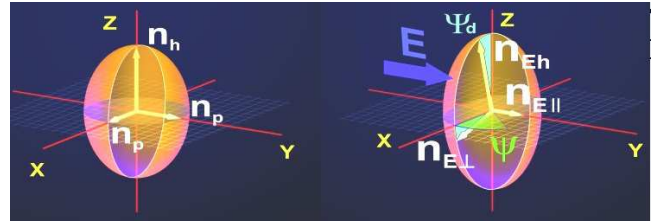}}
\caption{%
Ellipsoids of effective refractive indices of a short-pitch VADHFLC
cell. Left: at $\vc{E}=0$, the field-free effective
ellipsoid is uniaxially anisotropic with the optical axis parallel to
the helix axis. Right: applying an in-plane electric field $\vc{E}\parallel\uvc{y}$, makes
the optical anisotropy biaxial  with the two optical axes
rotated by the angle $\Psi_d\propto E$ about 
the electric field vector $\vc{E}$.~\cite{Kiselev:pre:2013,Kisel:pre:2011}
}
\label{fig:ellipsoid}
\end{figure}

For the pure phase modulation, where effects of diffraction, light scattering and
polarization plane rotation of the incident light are suppressed
in the visible spectral range, 
the selective reflection band of
the VADHFLC should be in the UV region. 
This can be achived when $p_0\ll\lambda$. 
For this case, the field-free ($\vc{E}=0$)
refractive index ellipsoid of the VADHFLC 
with subwavelength pitch is uniaxial (see Fig.~\ref{fig:ellipsoid}(left)), 
whereas applying the electric
field perpendicular to the helical axis induces optical biaxiality 
and rotation of the optical axes (see Fig.~\ref{fig:ellipsoid}(right)).~\cite{Kiselev:pre:2013,Kisel:pre:2011}

At $E\ne 0$,
the refractive indices
perpendicular 
($n_{E\perp}$)
and parallel ($n_{E \parallel}$)
to the electric field 
that govern
propagation of normally incident light beams 
can be written as folows~\cite{Kiselev:pre:2013}:
\begin{align}
&
  \label{eq:n_perp}
  n_{E \perp}/n_p=1+\frac{\epsilon_e-\epsilon_{\perp}}{\epsilon_e+\epsilon_{\perp}}
\left[
\frac{\epsilon_0 \chi_{G}}{P_s}
\right]^2 E^2,
\\
&
\label{eq:n_par}
  n_{E \parallel}/n_p=1-\frac{\epsilon_e-\epsilon_{\perp}}{\epsilon_e+\epsilon_{\perp}}
\left[
\frac{\epsilon_0 \chi_{G}}{P_s}
\right]^2 E^2,
\\
&
\label{eq:n_p}
n_{p}=\sqrt{(\epsilon_e+\epsilon_{\perp})/2},
\:
\epsilon_e=\frac{\epsilon_{\parallel}\epsilon_{\perp}}{\epsilon_{\perp}\sin^2\theta+\epsilon_{\parallel}\cos^2\theta},
\end{align}
where
$\epsilon_{\parallel}$ ($\epsilon_{\perp}$)
is the high frequency dielectric
constant measured parallel (perpendicular) to 
the FLC director;
$\theta$ is the smectic tilt angle;
$P_s$ is the spontaneous ferroelectic polarization
and
$\epsilon_0$
is the dielectric permittivity of free space.
The dielectric susceptibility of the Goldstone
mode, $\chi_G$, is given by~\cite{Urbanc:ferro:1991}
\begin{align}
  \label{eq:chi_G}
  \chi_{G}=\epsilon_0^{-1}\frac{\partial P}{\partial E}=
\frac{P_s^2}{2\epsilon_0 K q_0^2\sin^2\theta},
\end{align}
where $q_0=2\pi/p_0$ and $K$ is the effective twist elastic constant.

From Eqs.~\eqref{eq:n_perp}--~\eqref{eq:chi_G},
the field-induced in-plane birefringence
\begin{align}
&
  \label{eq:delta_n}
  \delta n_{i}=n_{E \perp}-n_{E \parallel}=
\notag
\\
&
2n_p
\frac{\epsilon_e-\epsilon_{\perp}}{\epsilon_e+\epsilon_{\perp}}
\left[
\frac{\epsilon_0 \chi_{G}}{P_s}
\right]^2 E^2=
K_{\mathrm{kerr}}\lambda E^2
\end{align}
can be expressed in terms of the Kerr constant
given by
\begin{align}
  \label{eq:K_kerr}
  K_{\mathrm{kerr}}=
\frac{n_p}{\lambda}
\frac{\epsilon_e-\epsilon_{\perp}}{\epsilon_e+\epsilon_{\perp}}
\frac{P_s^2 p_0^4}{32\pi^2 K^2\sin^4\theta}.
\end{align}
From Eq.~\eqref{eq:K_kerr} it is clear that 
an increase in $P_s$ and $p_0$ will enhance the Kerr constant $K_{\mathrm{kerr}}$. 
There are, however, certain limitations on these parameters:
(a)~the pitch $p_0$ should be sufficiently small so
as to avoid the effects of selective reflection band mentioned above; 
(b)~an increase in $P_s$ of a chosen FLC  mixture 
requires high concentrations of chiral molecules 
that will increase the rotational viscosity of the mixture
$\gamma_{\phi}$ 
thus affecting the EO response time.~\cite{Barnik:mclc:1987,Pozhidaev:jetp:1988}  
So, there is a tradeoff between
high value of $K_{\mathrm{kerr}}$ and fast EO response.
In our experiments, the optimal relationship
between the parameters 
$\gamma_{\phi}$, $\theta$, $K$, $P_s$ and $p_0$ 
leading to a drastic increase of $K_{\mathrm{kerr}}$ 
has been achieved with a proper choice of 
specially selected chemical
structures of the FLC mixture components and their
concentrations.

The method of 
mixing achiral smectic C and chiral
nonmesogenic compound~\cite{Kuczynski:chempl:1980} 
was employed to elaborate
a FLC mixture  that meets the above 
constraints.
A peculiarity of the mixture 
is that it contains two chiral nonmesogenic
compounds, which induce in achiral smectic C matrix
spontaneous polarization of the same sign, 
whereas their optical twisting powers are
opposite in sign.~\cite{Pozhid:lc:1989} 
This enables the relation between $P_s$ and $p_0$ 
to be fine tuned while the value of $P_s$ increases.

\begin{figure}[!tbh]
\centering
\resizebox{80mm}{!}{\includegraphics*{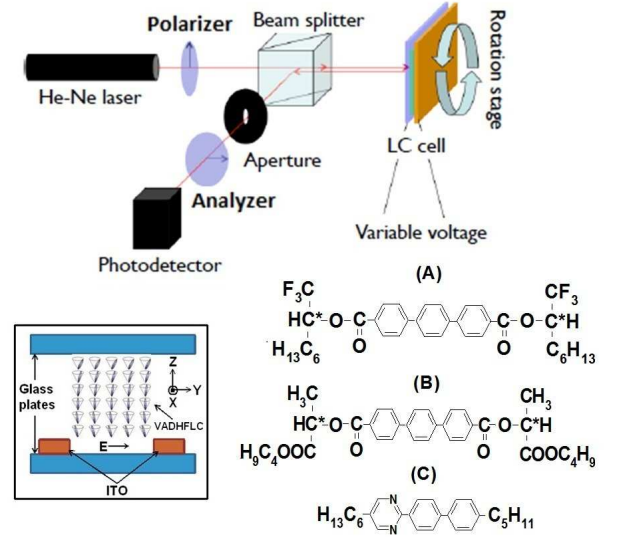}}
\caption{%
Top: Experimental set-up for reflectance measurements. Glan prisms
were used as both polarizer and analyzer. 
Bottom (left): VADHFLC cell
with ITO electrodes (the width is $20$~$\mu$m and the inter-electrode gap is
100 $\mu$m) and the axes of the coordinate system shown in Fig.~\ref{fig:ellipsoid}; 
Bottom (A-C): Chemical structures of compounds used to prepare the FLC-618. 
}
\label{fig:setup}
\end{figure}

One of these compounds is a fluorinated derivative of
the p-terphenyldicarboxylic acid that exhibit extremely high
twisting power~\cite{Pozhidaev:mclc:2009} (see compound A in Fig.~\ref{fig:setup}).
Another chiral compound is a lactate derivative of the p-terphenyldicarboxylic acid shown
as compound B in Fig.~\ref{fig:setup}.
Achiral smectic C matrix of the mixture
is the biphenylpirimidine indicated as compound C in Fig.~\ref{fig:setup}.

The eutectic mixture of these compounds
corresponds to the following weight concentrations:
39\% of compound A, 9\% of compound B, 
and 52\% of compound C. 
The mixture is named FLC-618 and is
characterized by
$p_0=175$~nm and
$P_s=200$~nC/cm$^2$
(at temperature $T=22$\degc).
The phase transitions sequence of this
FLC at heating from preliminary obtained solid crystalline state is
$
\mathrm{Cr}\myarrow{+21}\mathrm{Sm}C^{\star}\myarrow{+100}\mathrm{Sm}A^{\star}\myarrow{+117} \mathrm{Iso},
$
whereas at cooling down from isotropic phase crystallization occurs around +6\degc.

\begin{figure}[!tbh]
\centering
\resizebox{80mm}{!}{\includegraphics*{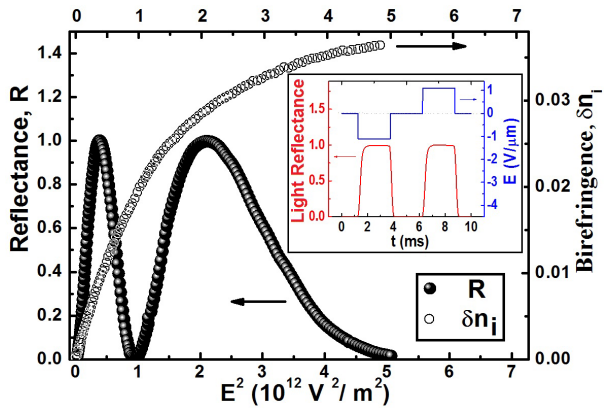}}
\caption{%
Dependence of the light reflectance of a 18 $\mu$m thick VADHFLC cell on
square of electric field $E^2$ (filled circles). Measurements were carried out in the
reflective mode with crossed polarizer and analyzer at temperature
55\degc, wavelength $\lambda=632.8$ nm, 
electrooptical response frequency 500Hz
and $\Psi=45$\dega. 
Electric field-induced birefringence $\delta n_{i}$ (open circles) was
evaluated as a function of $E^2$ by using Eq.~\eqref{eq:R}.
Insert:
Electrooptical response of the VADHFLC cell (red curve at the
bottom) under the applied alternating signal (blue curve on the top).
}
\label{fig:biaxiality}
\end{figure}

Electro-optical studies 
have been carried out in the reflective geometry
shown in Fig.~\ref{fig:setup} with a VADHFLC layer of
thickness $d_{\mathrm{FLC}} = 18$~$\mu$m 
placed between either crossed or parallel
polarizers at $\Psi=45$\dega
and $\Psi=0$,
where
$\Psi$ angle is the angle between the polarization plane of
incident light and the $X$ axis normal to the electric field 
(see Fig.~\ref{fig:ellipsoid}). 
A helium-neon laser with wavelength either
632.8 nm or 543 nm was used as a source of light. 
The EO response of a VADHFLC cell at 
$E = 1$~V/$\mu$m shown in the insert of Fig.~\ref{fig:biaxiality} 
confirms good optical quality (the contrast ratio is
better than 1000:1) and response time around 300 $\mu$s.

When polarizer and analyzer are parallel
and $E=0$,
the intensity of light,
$I_{\parallel}^{E=0}$,
that after all reflections
(from dielectric surfaces of the cell, beam splitter and
Glan prisms used as the polarizers)
is collected by the photodetector,
is proportional to the laser beam intensity, $I_0$: 
$I_{\parallel}^{E=0}= r I_{0}$. 
In the case of crossed polarizers with $\Psi\ne 0$ and $E\ne 0$, 
the light intensity collected by the photodetector 
$I_{\perp}^{E\ne 0}$ can be conveniently
expressed in terms of the normalized reflectance given by
\begin{align}
  \label{eq:R}
  R =
I_{\perp}^{E\ne 0}/I_{\parallel}^{E=0}=
\sin^2(2\Psi)\sin^2\frac{ 2 \pi\delta n_i d_{\mathrm{FLC}}}{\lambda},
\end{align}
where $\delta n_i$ plays the role of
field-induced in-plane birefringence for a light beam
propagating along the normal to the cell (the $Z$ axis).

Formula~\eqref{eq:R} can now be used to evaluate
$\delta n_i$
from the measured reflectance $R$.
Typical results for the electric field dependencies
of $R$ and $\delta n_i$ are presented in Fig.~\ref{fig:biaxiality}.
In the reflective mode, the phase retardation resulted from 
the electrically induced biaxiality can be expressed in terms of the
Kerr constant as follows
\begin{align}
  \label{eq:dphi}
  \Delta\Phi_{\ind{ret}}=\frac{4 \pi\delta n_i d_{FLC}}{\lambda}=
4\pi K_{\mathrm{kerr}} E^2 d_{\mathrm{FLC}}.
\end{align}

Referring to Fig.~\ref{fig:biaxiality}, it is clear that, for $\Psi=45$\dega,
$\Delta\Phi_{\ind{ret}}=4 \pi$ at $E\approx 2.3$~V/$\mu$m.
It can also be seen that, 
owing to rapidly growing higher order nonlinearities,
the Kerr-like dependence~\eqref{eq:delta_n} is no longer
valid at the phase retardation above 2$\pi$
($E\ge 1$V/$\mu$m).

\begin{figure}[!tbh]
\centering
\resizebox{80mm}{!}{\includegraphics*{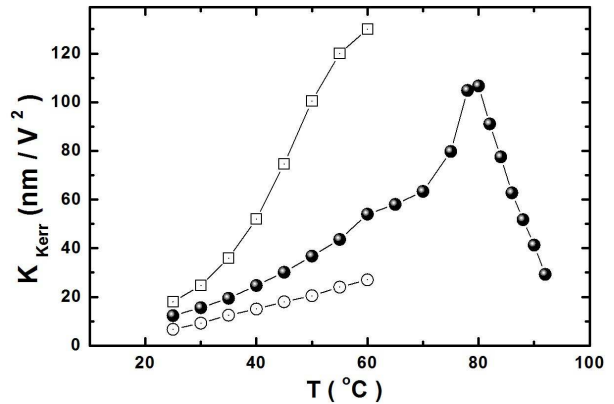}}
\caption{%
Temperature dependencies of Kerr constant  
measured in a FLC-618 cell at
$\lambda=632.8$~nm (filled circles) and at $\lambda=543$~nm 
(open squares).
Open circles represent the data measured in a FLC-587 cell at
$\lambda=543$~nm.~\cite{Kiselev:pre:2013}
}
\label{fig:Kerr}
\end{figure}

The curves representing
the Kerr constant, $K_{\mathrm{kerr}}$,
evaluated at different temperatures
are plotted in Fig.~\ref{fig:Kerr}.
It is seen that
$K_{\mathrm{kerr}}$ initially increases with
temperature, reaches the maxima 
and decreases at higher temperatures. 
This behavior of $K_{\mathrm{kerr}}$ can be explained by the temperature dependence of $P_s$ and
$p_0$ that enter Eq.~\eqref{eq:K_kerr}.
The largest value of $K_{\mathrm{kerr}}$
evaluated for our VADHFLC is 130~nm/V$^2$.

\begin{figure}[!tbh]
\centering
\resizebox{80mm}{!}{\includegraphics*{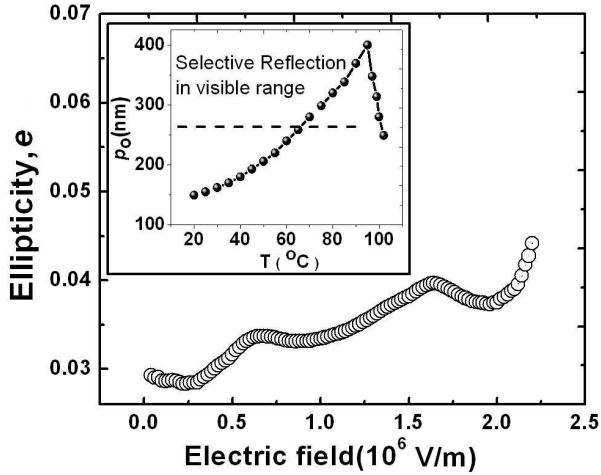}}
\caption{%
Ellipticity of light reflected from a $18$~$\mu$m thick
VADHFLC cell versus applied electric field measured 
at $\Psi=0$, $f=500$~Hz,
$T=55$\degc, and $\lambda=632.8$~nm. 
Insert: Temperature dependence of the helix pitch for FLC-618. 
Selective reflection is observed
when the pitch is above the dashed line.
}
\label{fig:ellipt}
\end{figure}

In order to demonstrate how relation~\eqref{eq:K_kerr} 
may guide the way to enhance 
the Kerr constant using classical methods of FLC material science,
we compare the results for two FLC mixtures:
the newly developed FLC-618
and the mixture FLC-587 described in.~\cite{Kiselev:pre:2013}
The parameters of FLC-587
at temperature 60\degc\  are: 
$\theta=35$\dega, $p_0 = 210$~nm, 
$P_s = 110$~nC/cm$^2$. 
Similarly, for FLC-618 at $T=60$\degc\, we have:
$\theta=33$\dega, $p_0 = 250$~nm
(see insert in Fig.~\ref{fig:ellipt}), 
$P_s = 160$~nC/cm$^2$.
 These
parameters combined with Eq.~\eqref{eq:K_kerr} 
can now be used to obtain a theoretical estimate
for the Kerr constants ratio 
$K_{\mathrm{kerr}}^{FLC-618}/K_{\mathrm{kerr}}^{FLC-587}\approx 5.2$.
This estimate agrees reasonably well
with the experimental value of this ratio which is about 4.8 
(see Fig.~\ref{fig:Kerr}). 

It should be emphasized that,
at $\Psi=45$\dega,
when the phase retardation~\eqref{eq:dphi} 
continuously changes between 0 and 4$\pi$, 
the ellipticity of reflected light significantly varies. 
This
was checked experimentally using the known
polarimetry method.~\cite{Kisel:pra:2008} 
In other words, 
the phase retardation arises from 
the electrically controlled birefringence~\eqref{eq:delta_n}
so that both amplitude and phase
modulation of light are observed. 
This geometry of VADHFLC is
more appropriate for display devices 
and, as compared to other alternatives,
has the important advantage of large $K_{\mathrm{kerr}}$.

At $\Psi=0$,
the polarization plane of incident light is parallel 
to the plane where the optical axes lie 
(the $XZ$ plane in Fig.~\ref{fig:ellipsoid}) 
and the incident light  beam does not split into ordinary 
and extraordinary rays.
In this case, similar to B-effect in nematic LCs,
the phase shift $\Delta\Phi_{\perp}$
of incident light  occurs solely due to 
changes in the refractive index $n_{E \perp}$.
This is the regime of  pure phase modulation. 
It is not difficult to see that
the phase
shift at $\Psi=0$ is equal to one-half  
the phase retardation
at $\Psi=45$\dega:
\begin{align}
  \label{eq:phase-shift}
  \Delta\Phi_{\perp}=
\frac{4 \pi (n_{E \perp}-n_{p}) d_{FLC}}{\lambda}=
\frac{1}{2} \Delta\Phi_{\ind{ret}}.
\end{align}
So, 
at $\Psi=0$ and $E = 2.3$~V/$\mu$m, 
we have the 2$\pi$ phase shift. 

Referring to Fig.~\ref{fig:ellipt},
when the phase shift $\Delta\Phi_{\perp}$ changes 
between 0  and 2$\pi$,
variations of the ellipticity are ranged 
from $0.027$ to $0.045$. 
Experimental imperfections 
and small fluctuations in orientation of the optical axes
that are neglected in the theory~\cite{Kiselev:pre:2013}
can be attributed for these variations. 
For practical applications, this can be regarded as
a sub-kilohertz pure phase modulation of light with the
conserved ellipticity.

In conclusion, 
based on the theoretical analysis and material optimization, 
we revealed an efficient strategy to enhance the Kerr effect
in VADHFLCs and have achieved the large value of the Kerr constant,
$K_{\mathrm{kerr}}\approx 130$~nm/V$^2$.
Further improvements in material optimization for $P_s$ and $p_0$ of the FLC
material are possible to accomplish the pure 2$\pi$ phase modulation at
room temperature. However, these optimizations will impose tight
constrains on the selective reflection limited to UV region and the
optimized rotational viscosity of the FLC mixture. The proposed
VADHFLC systems with a continuous, hysteresis-free 2$\pi$ phase modulation
at sub kHz frequencies could find application in many modern photonic
and display devices demanding fast phase only modulation at 
low electric fields.

This work is supported by the HKUST grants CERG 612310
and RGC 614410 and
by RFBR grants 13-02-00598\_A, 1302-90487\_Ukr\_f\_a.


%

\end{document}